\documentclass[aps,prx,superscriptaddress,amsmath,amssymb,twocolumn,showpacs,floatfix,reprint]{revtex4-2}
\usepackage{multirow}
\usepackage[dvipsnames]{xcolor}
\usepackage[utf8]{inputenc}
\usepackage{braket}
\usepackage{graphicx}
\usepackage{xr}
\usepackage[colorlinks=true, urlcolor=blue, linkcolor=blue, citecolor=blue, pdftex]{hyperref}
\usepackage[capitalise]{cleveref}

\usepackage{changes}

\begin{document}
\title{Superconductivity in the $t$-$t'$ Hubbard Model from \\Symmetry-Preserving Neural-Network Quantum States}

\author{Riccardo Rende}
\thanks{These authors contributed equally. Correspondence should be addressed to rrende@flatironinstitute.org and luciano.viteritti@epfl.ch}
\affiliation{Center for Computational Quantum Physics, Flatiron Institute, 162 5th Avenue, New York, NY 10010}

\author{Luciano Loris Viteritti}
\thanks{These authors contributed equally. Correspondence should be addressed to rrende@flatironinstitute.org and luciano.viteritti@epfl.ch}
\affiliation{Institute of Physics, \'{E}cole Polytechnique F\'{e}d\'{e}rale de Lausanne (EPFL), CH-1015 Lausanne, Switzerland}

\author{Antoine Georges}
\affiliation{Coll{\`e}ge de France, 11 place Marcelin Berthelot, 75005 Paris, France}
\affiliation{Center for Computational Quantum Physics, Flatiron Institute, 162 5th Avenue, New York, NY 10010}
\affiliation{CPHT, CNRS, {\'E}cole Polytechnique, IP Paris, F-91128 Palaiseau, France}
\affiliation{DQMP, Universit{\'e} de Gen{\`e}ve, 24 quai Ernest Ansermet, CH-1211 Gen{\`e}ve, Suisse}

\date{\today}

\begin{abstract}
Despite its fundamental importance in the theory of strongly correlated electrons, the nature of the ground state of the two-dimensional doped Hubbard model remains intensely debated. Variational approaches provide a powerful route to this problem, but their conclusions can depend sensitively on the chosen wave-function parameterization, the mean-field initialization, or the pinning fields used to guide the optimization, as well as on boundary conditions. This can favor one type of symmetry breaking over another, making it difficult to distinguish the genuine interplay of intertwined or competing orders from biases induced  by the variational parameterization. Here, we introduce the Symmetry-Preserving Backflow Pairing (SBP) ansatz, a neural-network wave function that respects translational symmetry by construction and thereby avoids these broken-symmetry minima. The SBP ansatz reaches state-of-the-art variational energies for the $t$-$t'$ Hubbard model on lattices up to $24\times24$ with $504$ electrons, below those of competing pure stripe solutions. By extrapolating to the thermodynamic limit, we find robust evidence for $d$-wave superconducting order, resolving a long-standing question about the $1/8$-doped model at $t'/t=-0.2$ and $U/t=8.0$. Built on general principles of symmetry and locality, the SBP wave function provides a broadly applicable variational representation for challenging interacting fermionic systems.
\end{abstract}

\maketitle

\section{Introduction}
The two-dimensional Hubbard model is one of the simplest models of strongly correlated electrons, and it is widely regarded as a minimal model that qualitatively captures key physical aspects of the copper-oxide superconductors ~\cite{hubbard1963,Arovas2022,Qin2022}. Despite its simple form, its phase diagram at finite doping is still not settled~\cite{gull2015}. Antiferromagnetism, charge and spin modulations, and unconventional pairing arise from the same microscopic interaction; the corresponding states are very close in energy, and these intertwined orders~\cite{rmp2015_intertwined} can coexist~\cite{shiwei2024science,roth2025}. As a result, different numerical methods have  reached different conclusions about which state has the lowest energy~\cite{white1989, tocchio2016, shiwei2016qmc, simkovic2024, sorella2023, solving2025, roth2025, Marino_2022}.

Progress on this problem has been substantial~\cite{gull2015}, yet a striking feature is that different conclusions are reached not only by different methods but also within the same numerical approach~\cite{huang2018, science2019dmrg, npj2026}. A recent large-scale study combining density-matrix renormalization group (DMRG) with constrained-path auxiliary-field quantum Monte Carlo (AFQMC) reports that, when the next-nearest-neighbor hopping $t'$ is negative, superconductivity coexists with partially filled stripes~\cite{shiwei2024science}, whereas a separate DMRG study finds no superconductivity for $t'<0$ and remains inconclusive about its presence for $t'>0$~\cite{npj2026}. A similar situation arises for Neural-Network Quantum States (NQS)~\cite{carleo2017}. A wave function built from backflow determinants finds only pure stripe order~\cite{solving2025,gu_pareto_2026}, while a Pfaffian-based wave function reports clear signatures of superconductivity for $t'<0$ and even evidence for superconductivity in the pure Hubbard model at $t'=0$~\cite{roth2025}. 
A likely origin of these discrepancies is that each numerical method carries its own bias, mainly through the boundary conditions and the lattice geometries in tensor network-based approaches~\cite{huang2018, science2019dmrg, liu2025, npj2026} and through the initial mean-field state used to initialize the optimization~\cite{roth2025} or the specific functional form of the variational state in NQS~\cite{viteritti2026bias}.

In this work, we focus on the paradigmatic $t$-$t'$ Hubbard model, whose Hamiltonian reads
\begin{equation}
\hat{H} = -t \sum_{\langle ij \rangle, \sigma} \hat{c}^{\dagger}_{i\sigma} \hat{c}_{j\sigma} - t' \sum_{\langle\langle ij \rangle\rangle, \sigma} \hat{c}^{\dagger}_{i\sigma} \hat{c}_{j\sigma} + U \sum_{i} \hat{n}_{i\uparrow} \hat{n}_{i\downarrow} \ ,
\label{eq:Hubbard}
\end{equation}
where $i$ and $j$ label the sites of the lattice, $\hat{c}^{\dagger}_{i\sigma}$ ($\hat{c}_{i\sigma}$) creates (annihilates) an electron with spin $\sigma=\{\uparrow,\downarrow\}$ on site $i$, and $\hat{n}_{i\sigma}=\hat{c}^{\dagger}_{i\sigma}\hat{c}_{i\sigma}$ is the number operator. The sums $\langle ij \rangle$ and $\langle\langle ij \rangle\rangle$ run over nearest- and next-nearest-neighbor pairs, with hopping amplitudes $t$ and $t'$, while $U$ is the on-site repulsion. 
We set $t=1.0$ as the unit of energy and take $t'=-0.2$ and $U=8.0$ 
in order to compare with previous works~\cite{shiwei2024science,npj2026}. 
These values are in the range relevant to cuprate superconductors, with 
$|t'/t|$ varying from $\sim 0.15$ to $\sim 0.4$ depending on the compound
~\cite{andersen1995, hirayama2018, schmid2023}.
We study $L\times L$ square lattices ($N=L^2$) with periodic boundary conditions and a number of electrons $N_e=N(1-\delta)$ chosen such that the doping level is 
fixed to $\delta=1/8$. 

The most common fermionic NQS parametrizations are based on backflow determinants~\cite{luo2019}, and they often converge to states that break the symmetries of the Hamiltonian, even in the presence of periodic boundary conditions. In the Hubbard model, these solutions typically display charge and spin stripe patterns that break translational symmetry~\cite{solving2025,gu_pareto_2026,zhou2024prb,hiddenfermion,vicentinihubbard,zhou2026,Liang_2026,ido2018, darmawan2018}.

In this work, we first demonstrate this behavior explicitly: when the variational manifold allows broken translational symmetry solutions, the optimization readily converges to several distinct pure stripe states with competitive variational energies. In the $t$-$t'$ Hubbard model [see \cref{eq:Hubbard}], this is a serious limitation because broken-symmetry solutions can display physical properties that differ markedly from those of the symmetric state, typically suppressing pairing correlations and leading to states with weak or absent superconducting order - see Ref.~\cite{viteritti2026bias} for a detailed discussion. We expect that a related mechanism may also affect DMRG and other related tensor-network approaches, where cylindrical geometries and open boundary conditions naturally favor the stabilization of pure stripe patterns~\cite{liu2025, zheng2017}. In addition, such pure stripe states may be represented more efficiently at finite bond dimension than superconducting states, an interpretation consistent with Ref.~\cite{npj2026}, where superconducting correlations are found to be underestimated at finite bond dimension.

On a finite cluster with periodic boundary conditions, a broken-symmetry state can be an exact ground state only in the presence of degeneracy, whereas when the Hamiltonian has symmetries, its exact eigenstates can always be chosen to transform according to definite irreducible representations of the symmetry group~\cite{wigner1959group}. Targeting the symmetric state directly, however, is not straightforward, since constructing expressive variational states that preserve translational symmetry by construction can be more challenging. Indeed, several works have shown that the most accurate variational results are often obtained by first allowing the wave function to break the symmetries of the Hamiltonian and then restoring them a posteriori through quantum-number projection~\cite{reh2023, tahara2008, nomura2021}. In the Hubbard model, however, the presence of competing pure stripe states makes this strategy effective only on relatively small system sizes~\cite{viteritti2026bias}, as discussed below.

\begin{figure}[t]
  \centering
  \includegraphics[width=\columnwidth]{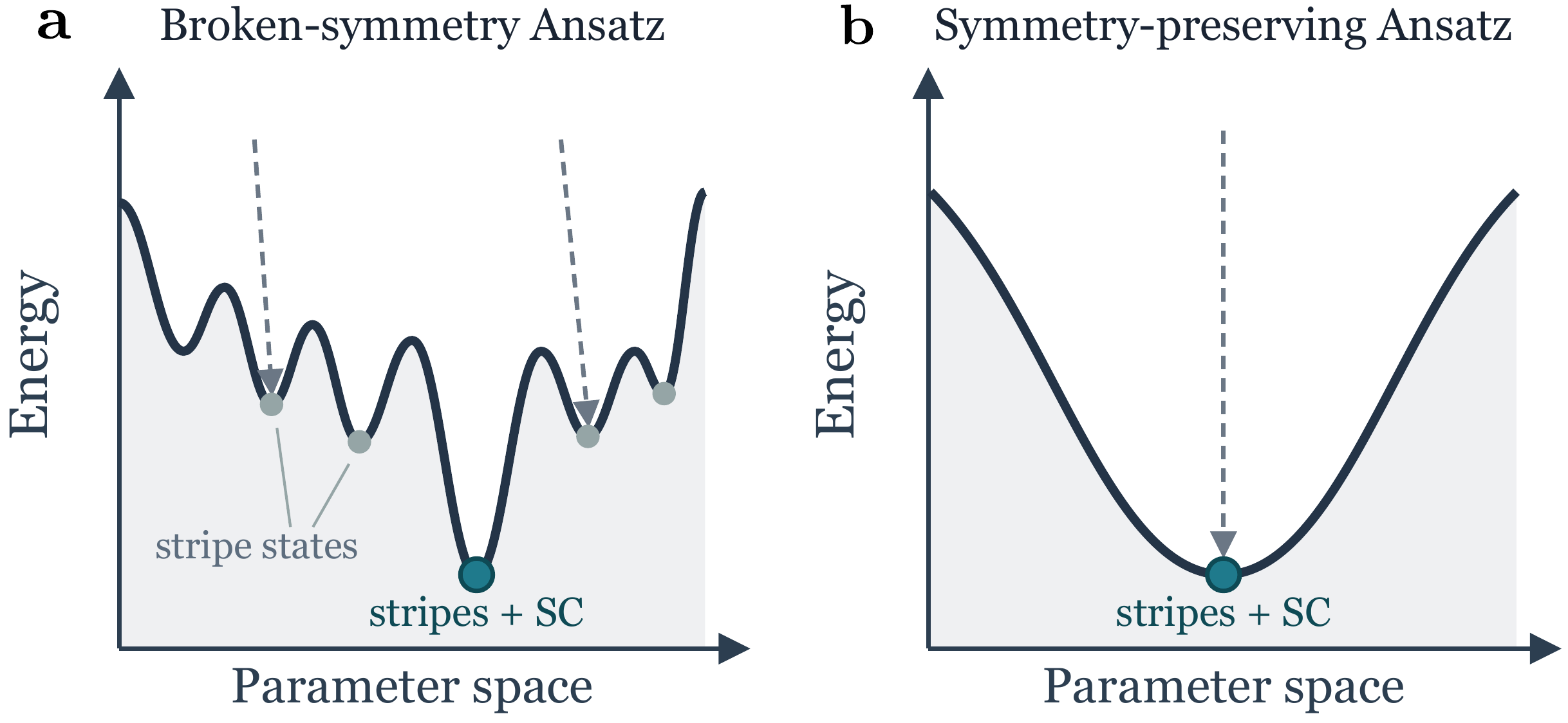}
  \caption{Pictorial representation of the variational energy landscape in the 
  relevant basin of attraction, as a function of the wave-function parameters (see main text). \textbf{Panel (a)}: With an ansatz that breaks the symmetries of the Hamiltonian the landscape is rugged: the optimization is easily trapped in local minima corresponding to broken-symmetry, pure stripe states (gray dots), and tends to miss the lower-lying minimum in which stripe and superconducting order can coexist (teal). \textbf{Panel (b)}: The proposed symmetry-preserving ansatz reshapes the landscape into an effectively ``convex'' one, with a single minimum to which the optimization converges reliably.  
  }
  \label{fig:landscape}
\end{figure}

In order to overcome these problems, we propose the Symmetry-Preserving Backflow Pairing (SBP) wave function that is designed to respect the translational symmetry of the Hamiltonian and, unlike previous constructions~\cite{solving2025,gu_pareto_2026}, does not become trapped in the local minima associated with broken-symmetry solutions of spin or charge. By enforcing the relevant symmetries at the level of the variational parametrization, the ansatz appears to reshape the optimization landscape into an effectively ``convex'' one: starting from different random initializations, the optimization converges to different sets of parameters that nonetheless describe the same physical state, with compatible energy and correlation functions; see \cref{fig:landscape} for a pictorial illustration. This behavior is consistent with the overparametrized regime typical of deep learning, where distinct sets of parameters can represent the same function and the minima reached from different initializations are effectively equivalent~\cite{liu2022,garipov2018loss,draxler2018barriers}. In practice, this means that the basin of attraction of the minimum corresponding to the lowest variational energy is enlarged when considering a symmetric ansatz, making the landscape 
more convex in its vicinity - while other local minima outside this basin can of course 
still exist.

Our method reaches state-of-the-art variational energies, lower than those of competing pure stripe solutions on the same clusters, for lattices up to $24\times24$ and $504$ electrons. By extrapolating the superconducting order parameter to the thermodynamic limit, we find strong numerical evidence for robust $d$-wave superconductivity. 

These results address a fundamental long-standing question about the nature of the ground state of the $1/8$-doped model Hubbard model at $t'=-0.2$ and $U=8.0$, 
a parameter range relevant to cuprate superconductors. 

\section{Symmetry-Preserving Backflow Pairing} 
We introduce the Symmetry-Preserving Backflow Pairing (SBP) ansatz, a generalization of the pair-product wave function~\cite{nomura2017,tahara2008, ido2018, darmawan2018}. The variational state $\Psi_{\theta}(\boldsymbol{n})$, with $\theta$ a set of parameters, takes as input a physical configuration ${\boldsymbol{n}=(n_{1\uparrow},\ldots,n_{N\uparrow},n_{1\downarrow},\ldots,n_{N\downarrow})}$, encoded as a binary string of occupations $n_{i\sigma}\in\{0,1\}$. The wave function is defined as
\begin{equation}
\Psi_{\theta}(\boldsymbol{n})=\langle\boldsymbol{n}|\left(\sum_{i,j=1}^{N}f_{ij}(\boldsymbol{n})\,c^{\dagger}_{i\uparrow}c^{\dagger}_{j\downarrow}\right)^{N_e/2}|0\rangle,
\label{eq:ansatz}
\end{equation}
where $N_e$ is the number of electrons and $f(\boldsymbol{n})$ is a configuration-dependent $N\times N$ pairing matrix generated by the neural network. Since the backflow pairing orbitals only connect up and down spins [see \cref{{eq:ansatz}}], the amplitude can be evaluated as a determinant instead of a Pfaffian~\cite{tahara2008,bouchaud1988pair}:
\begin{equation}
\Psi_{\theta}(\boldsymbol{n})=\det[\,\boldsymbol{n}_{\uparrow}\star f(\boldsymbol{n})\star\boldsymbol{n}_{\downarrow}\,],
\end{equation}
where $\boldsymbol{n}_{\uparrow}\star f(\boldsymbol{n})\star\boldsymbol{n}_{\downarrow}$ denotes the $N_e/2\times N_e/2$ submatrix of $f(\boldsymbol{n})$ obtained by retaining only the rows and columns corresponding to occupied up- and down-spin sites, respectively. This formulation allows the spatial symmetries of the Hamiltonian to be incorporated directly through the parametrization of $f(\boldsymbol{n})$. 

In this work, we introduce the following backflow form for the pairing matrix~\cite{loehr2025, chen2025, viteritti2026bias}:
\begin{equation}
        f_{ij}(\boldsymbol{n}) = \frac{e^{-\lambda d(i, j)}}{\sum_{i'j'}e^{-\lambda d(i', j')}} \sum_{\alpha=1}^{d} W_{i-j,\,\alpha} \, y^{\uparrow}_{i\alpha}(\boldsymbol{n}) \, y^{\downarrow}_{j\alpha}(\boldsymbol{n}) \ ,
    \label{eq:backflow}
\end{equation}
where the backflow transformation produces the $d$-dimensional vectors $y^{\sigma}_{i}(\boldsymbol{n})$, which provide an abstract, configuration-dependent representation of the input configuration $\boldsymbol{n}$. These vectors are generated by a translationally equivariant transformer-based architecture~\cite{viteritti2023prl, viteritti2025prb}, additional details are given in the \textit{Methods}. The matrix of $W_{i-j,\alpha}$ depends only on the relative displacement $i-j$ to preserve translational symmetry: this results in $N \times d$ independent trainable parameters.
For square geometries, the $C_{4v}$ point-group symmetry can also be incorporated explicitly into the variational state (see \textit{Methods}). Since this group has only eight elements, however, we restore it a posteriori by quantum-number projection~\cite{tahara2008, rende2024stochastic}.
\begin{figure}[t]
  \centering
  \includegraphics[width=\columnwidth]{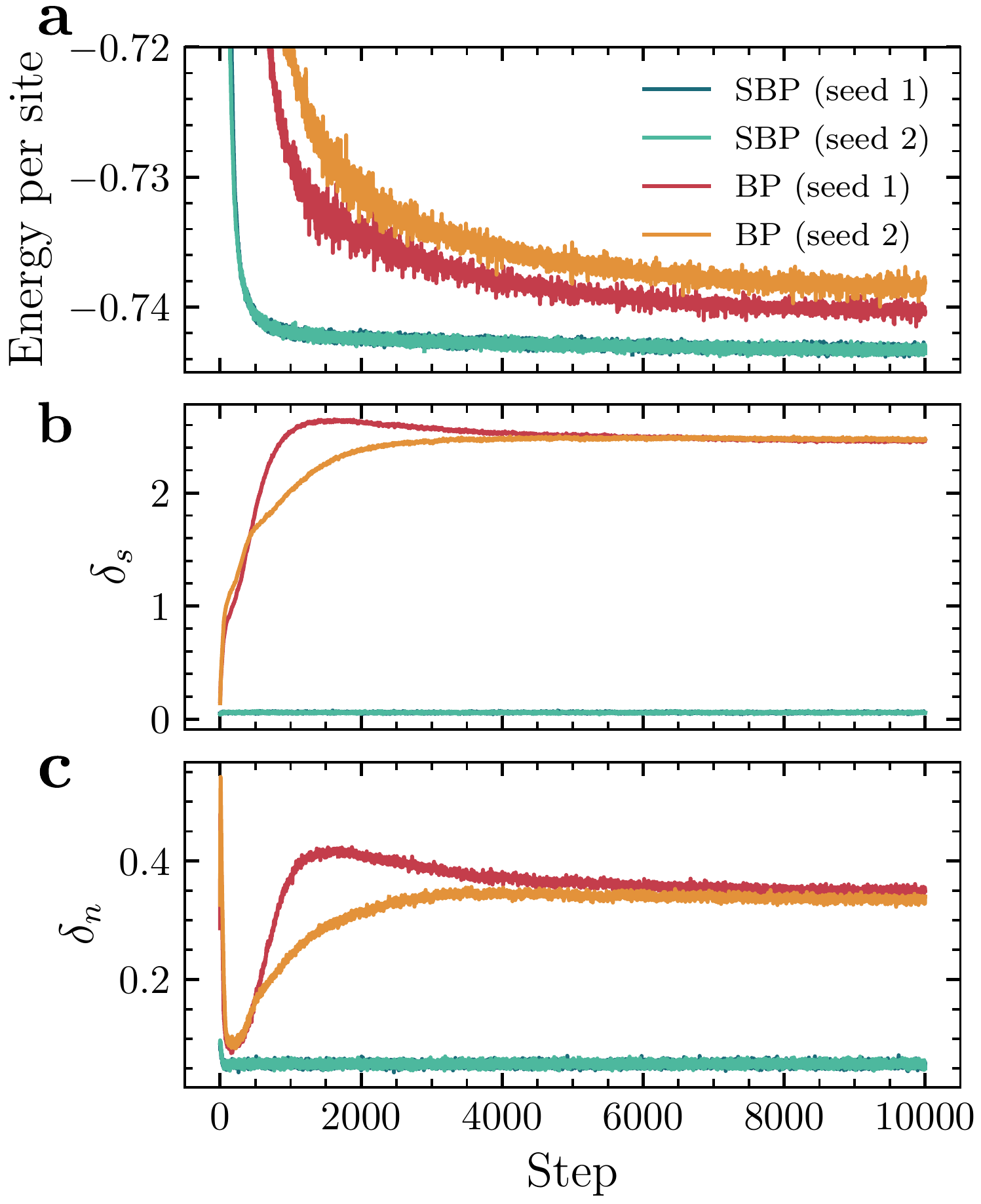}
    \caption{Optimization of the SBP wave function [see \cref{eq:ansatz}] and of the BP wave function [see \cref{eq:BP_ansatz}] for two different random seeds, for the $t$-$t'$ Hubbard model [see \cref{eq:Hubbard}] on a $12\times12$ lattice at $t'=-0.2$, $U=8.0$, and doping $\delta=1/8$. \textbf{Panel (a)}: Variational energy per site as a function of the optimization step. \textbf{Panel (b) and (c)}: Spin and charge symmetry breaking parameters $\delta_s$ and $\delta_n$ [see \cref{eq:order}] as a function of the optimization steps.}
  \label{fig:optimization}
\end{figure}
The prefactor $e^{-\lambda d(i,j)}/\sum_{i'j'} e^{-\lambda d(i',j')}$ in \cref{eq:backflow} introduces a spatial decay in the pairing matrix. This choice is motivated by the physical interpretation of $f_{ij}(\boldsymbol{n})$ as the amplitude for creating a pair of electrons on sites $i$ and $j$, with spins $\uparrow$ and $\downarrow$, respectively. For a local Hamiltonian, such amplitudes are expected to decay with the distance between the two sites, independently of the configuration $\boldsymbol{n}$. 
We emphasize that the locality of the pairing amplitude $f_{ij}$ does not imply short-range pairing correlations, as illustrated for example by the attractive Hubbard model, where a local pairing amplitude gives rise to off-diagonal long-range order. 
Here $d(i,j)$ is the Euclidean distance between sites $i$ and $j$ on the lattice, and $\lambda>0$ is a trainable parameter controlling the decay length. The denominator normalizes the weights over all pairs. In this way, the ansatz suppresses matrix elements between distant sites of the pairing matrix while allowing the optimization to determine the appropriate decay scale. This bias becomes important for large lattices; it stabilizes the optimization, speeds up convergence, and drives it to lower energy solutions. A related spatial bias has originally been used in the transformer attention mechanism to obtain stable and accurate optimization on large spin systems on lattice~\cite{viteritti2026approaching} and more recently in electronic systems in continuous space~\cite{gaggioli2026}.

\section{Symmetry breaking and local minima}
In this Section, we provide evidence for the presence of distinct local minima in the variational landscape when using variational wave functions that explicitly break lattice symmetries. To this end, we introduce a parametrization of the pairing matrix in \cref{eq:ansatz} that does not enforce translational symmetry. Following the construction of the backflow Pfaffian output layer introduced in Refs.~\cite{chen2025, loehr2025, viteritti2026bias}, we define
\begin{equation}
    \label{eq:BP_ansatz}
    f_{ij}(\boldsymbol{n})
    =\sum_{\alpha=1}^{d} \Phi^{\uparrow} _{i\alpha}(\boldsymbol{n}) \  \Phi^{\downarrow}_{j\alpha}(\boldsymbol{n}) .
\end{equation}
where $\Phi^{\sigma}_{i\alpha}$ are backflow orbitals obtained through a site-resolved linear mapping of the transformer outputs, namely ${\Phi^{\sigma}_{i\alpha}(\boldsymbol{n}) = \sum_{\beta=1}^{d} W^{\sigma}_{i\alpha\beta}\ y^{\sigma}_{i\beta}}(\boldsymbol{n})$.
In the following, we refer to this state as the Backflow Pairing (BP) ansatz.

In \cref{fig:optimization}, we compare the optimization of the symmetry-preserving backflow-pairing state, denoted as SBP and defined in \cref{eq:backflow}, with that of the BP state in \cref{eq:BP_ansatz}. \Cref{fig:optimization}(a) shows the variational energy per site as a function of the optimization step. The SBP state converges rapidly and reaches the lowest energy, whereas the two BP optimizations converge more slowly to higher energies. Crucially, the two SBP runs, initialized with different random seeds, converge to the same energy, whereas the final energies of the two BP runs depend on the random seed used to initialize the variational parameters. This shows that the BP optimization is trapped in distinct local minima, while the SBP ansatz reliably reaches the same solution regardless of initialization. We stress that our goal here is not to find the lowest-energy pure stripe state, but to demonstrate that explicitly breaking translational symmetry gives rise to multiple pure stripe solutions. The best pure stripe variational energies available in the literature~\cite{solving2025,gu_pareto_2026} are reported in \cref{table:energies}.

To quantify the emergence of symmetry breaking during the optimization, we introduce the quantities:
\begin{equation}
    \delta^2_n
    =
    \sum_{i=1}^{N}
    \left(
    \langle \hat{n}_i \rangle - n_0
    \right)^{2},
    \qquad
    \delta^2_s
    =
    \sum_{i=1}^{N}
    \langle \hat{S}^z_i \rangle^{2}.
    \label{eq:order}
\end{equation}
Here, $n_0=N_{\mathrm{e}}/N$ is the density. The quantities $\delta_n$ and $\delta_s$ measure the amplitude of spatial modulations in the local density and in the local magnetization, respectively. \Cref{fig:optimization}(b) and \cref{fig:optimization}(c) show their evolution during the optimization.
For the SBP ansatz, both remain very small and constant throughout the entire optimization, so the state remains translationally invariant with no spatial modulation of the local spin moment or charge density. For the BP, instead, $\delta_s$ and $\delta_n$ grow to finite values, showing that the optimization is driven toward broken-symmetry solutions with pure stripe ordering.

\begin{figure}[t]
  \centering
  \includegraphics[width=\columnwidth]{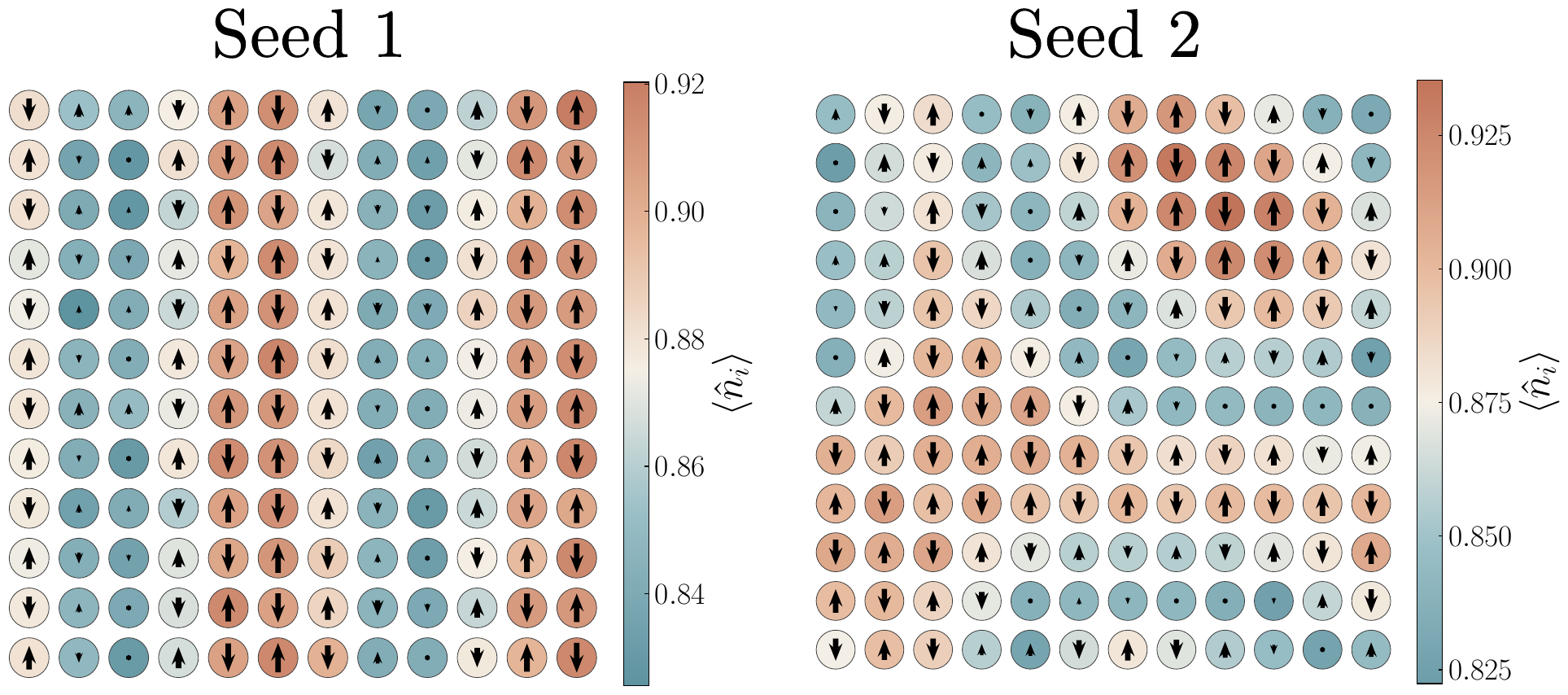}
  \caption{Local charge density $\langle \hat{n}_i \rangle$ and spin $\langle \hat{S}^z_i \rangle$ on each site of the $12\times12$ lattice, for the two BP solutions obtained from the two random seeds of \cref{fig:optimization}. The color of each site encodes the charge density $\langle \hat{n}_i \rangle$, as indicated by the colorbar, while the arrows show the local spin $\langle \hat{S}^z_i \rangle$.}
  \label{fig:patterns}
\end{figure}

To characterize the broken-symmetry solutions reached by the BP ansatz, in \cref{fig:patterns} we show the local charge density $\langle \hat{n}_i \rangle$ and the local magnetization $\langle \hat{S}^z_i \rangle$ for the two solutions obtained from the random seeds used in \cref{fig:optimization}. The color of each site encodes the charge density, as indicated by the color bar, while the arrows show the magnitude and sign of the local spin polarization. In both cases, the local observables are not uniform across the lattice but display clear spatial modulations. We emphasize that this behavior does not stem from a lack of expressivity of the BP parametrization: the state in \cref{eq:BP_ansatz} can also represent the translationally invariant solution of the SBP ansatz in \cref{eq:backflow}, but in practice, this solution is difficult to reach through unconstrained optimization.

Crucially, as shown in Ref.~\cite{viteritti2026bias}, these broken-symmetry solutions typically display weak or no superconductivity. One possible strategy to stabilize the superconducting solution would be to restore translational symmetry a posteriori by quantum-number projection~\cite{viteritti2026bias, tahara2008, nomura2021}. However, this becomes increasingly expensive with system size, as the projected wave function requires a sum over all translations, whose number grows with the number of lattice sites. The limitation may not be only computational: when the optimized state is strongly trapped in a broken-symmetry minimum, the translated configurations can have small overlap with the original one, which may hinder the optimization. Symmetry restoration is therefore effective on small clusters but impractical on large ones~\cite{viteritti2026bias}.

The occurrence of multiple stripe solutions is also observed in tensor-network approaches~\cite{liu2025}, where open boundary conditions favor stripe order, and pinning fields are often introduced to stabilize specific patterns~\cite{zheng2017, liu2025}. Similar behavior has also been reported in neural-network approaches; for example, Ref.~\cite{solving2025} compares different stripe solutions and measures the energy difference between horizontal and vertical stripes.
This contrasts with the SBP ansatz in \cref{eq:backflow}, for which the final solutions share the same energy and physical properties regardless of the random initialization of the parameters: in this case, the landscape appears to be effectively ``convex''~\cite{liu2022,garipov2018loss,draxler2018barriers} within the basin of attraction of the lowest energy variational state.

\section{Results} The SBP wave function achieves state-of-the-art variational results for the $t$-$t'$ Hubbard model. In \cref{table:energies}, we report the variational energies per site for $12\times12$, $16\times16$, $20\times20$, and $24\times24$ square lattices with periodic boundary conditions, together with the best available results in the literature. For a fair comparison, we include only energies obtained by direct variational optimization, excluding additional projector improvements such as Green Function Monte Carlo (GFMC)~\cite{trivedi1990, ceperley1995} or Lanczos steps~\cite{becca2015}. These techniques can be applied a posteriori to any optimized variational state and systematically lower the energy, but usually without inducing significant changes in its physical properties~\cite{ido2018}. On any system size, the SBP wave function reaches an energy that is lower than all previous variational results. 
Moreover, since the SBP variational state is translationally invariant, it is scalable across lattice sizes, and we can apply a transfer-learning procedure in which the wave function optimized on the $20\times20$ lattice is used as the starting point for the $24\times24$ lattice. In this way, the energy on the $24\times24$ lattice is obtained with $2000$ GPU hours on NVIDIA H200 nodes.

\begin{table}[b]
\centering
\begin{tabular}{c c c c c}
\hline\hline
Size & Ansatz & Energy & Variance & Ref. \\
\hline
\multirow{2}{*}{$12 \times 12$}
 & Transformer & -0.7414 & - & \cite{solving2025} \\
 & \textbf{SBP} & \textbf{-0.74425(1)} & \textbf{0.035(1)} & \textbf{This work} \\
\hline
\multirow{4}{*}{$16 \times 16$}
 & HFPS & -0.7335 & 0.07623 & \cite{Liang_2026} \\
 & Tensor-Backflow & -0.7360(1) & - & \cite{roth2025} \\
 & ACE & -0.7430 & - & \cite{gu_pareto_2026} \\
 & \textbf{SBP} & \textbf{-0.74411(1)} & \textbf{0.037(1)} & \textbf{This work} \\
\hline
\multirow{1}{*}{$20 \times 20$}
 & \textbf{SBP} & \textbf{-0.74382(1)} & \textbf{0.041(1)} & \textbf{This work} \\
\hline
\multirow{1}{*}{$24 \times 24$}
 & \textbf{SBP} & \textbf{-0.74397(1)} & \textbf{0.054(1)} & \textbf{This work} \\
\hline\hline
\end{tabular}
\caption{\label{table:energies} Ground-state variational energies and variances per site for different wave functions on the square lattice Hubbard model with PBC and parameters $t'=-0.2$, $U=8.0$ and doping $\delta = 1/8$.}
\end{table}

\begin{figure}[t]
  \centering
  \includegraphics[width=\columnwidth]{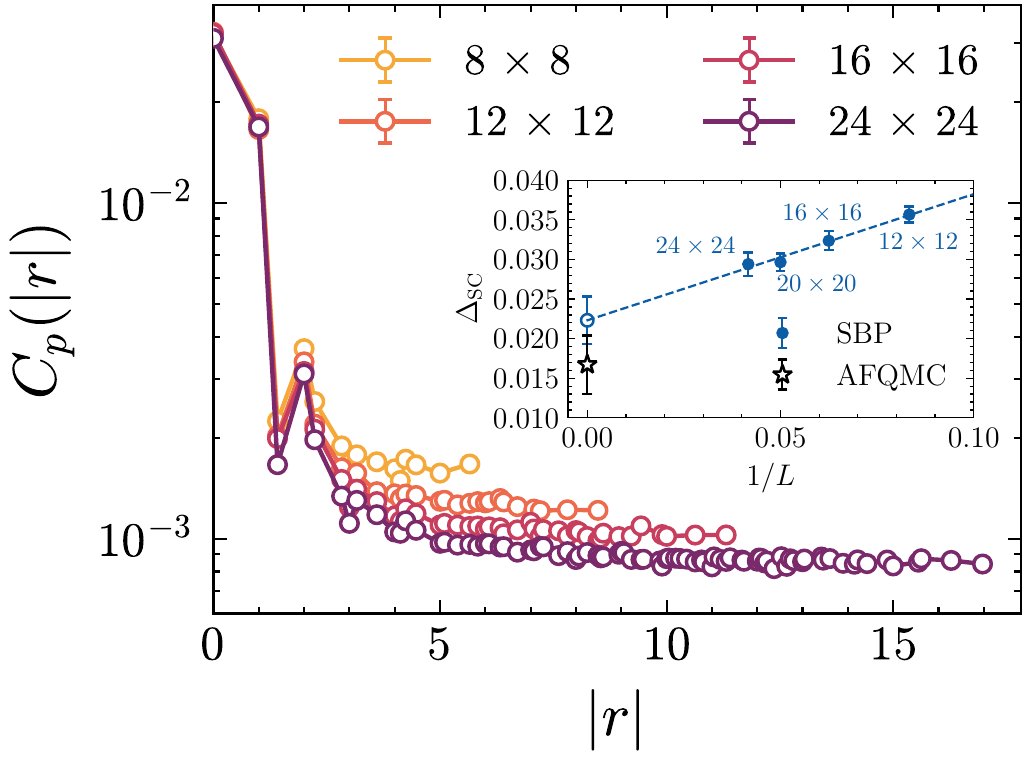}
  
    \caption{\label{fig:pairing_corr} $d$-wave pairing correlation function $C_p(|r|)$ [see \cref{eq:pairing_corr}] for the $t$-$t'$ Hubbard model with periodic boundary conditions, at $t'=-0.2$, $U=8.0$, and doping $\delta=1/8$, as a function of the distance $|r|$. \textbf{Inset}: $d$-wave order parameter $\Delta_{\mathrm{SC}}$ [see \cref{eq:pairing_op}] as a function of the inverse linear size $1/L$, together with the AFQMC results of Ref.~\cite{shiwei2024science}.
    }
\end{figure}

Beyond the variational energy, the physical properties of the $t$-$t'$ Hubbard model at finite doping remain under debate: antiferromagnetism, charge and spin modulations, 
and unconventional pairing all compete or coexist. Different numerical methods have reached different conclusions about whether superconductivity is present and whether it coexists with stripe order~\cite{zheng2017, ido2018, darmawan2018, science2019dmrg, qin2020, sorella2023, shiwei2024science, roth2025, viteritti2026bias}. Here, we provide strong numerical evidence of the presence of $d$-wave superconductivity in the $1/8$-doped model at $t'=-0.2$ and $U=8.0$, on large system sizes through the long distance behavior of the pairing correlation function~\cite{qin2020, roth2025, viteritti2026bias}:
\begin{equation}\label{eq:pairing_corr}
C_p(r)
=
\frac{1}{8}
\sum_{\eta,\eta'}
h_{0,\eta} h_{r,\eta'}
\left\langle
\hat{c}_{\eta,\downarrow}^{\dagger}
\hat{c}_{0,\uparrow}^{\dagger}
\hat{c}_{r,\uparrow}
\hat{c}_{r+\eta',\downarrow}
\right\rangle
-
\mathcal{N}_{0,r} \ ,
\end{equation}
the sum over $\eta$ runs over the four nearest neighbors of $r$, and $h_{r,\eta} = +1$ on horizontal bonds and $h_{r,\eta} = -1$ on vertical bonds, so that the pair carries symmetry $d_{x^2-y^2}$. The numerical prefactors in the definitions of $C_p(r)$ is chosen so that its
long-distance behavior can be directly compared with the standard convention used in Refs.~\cite{qin2020,shiwei2024science}. The term $\mathcal{N}_{0,r}$ subtracts the disconnected contribution~\cite{roth2025}. Refer to \textit{Methods} for additional details. 

A finite value of $C_p(r)$ at large distances signals superconducting $d$-wave order. In \cref{fig:pairing_corr} we show $C_p(r)$ as a function of distance $|r|$. After a fast decay at short distances, the correlation function saturates to a finite value as $|r|$ grows. The inset of \cref{fig:pairing_corr} shows the $d$-wave order parameter~\cite{ido2018}
\begin{equation}\label{eq:pairing_op}
    \Delta^2_{\mathrm{SC}}=\frac{1}{\mathcal{M}}\sum_{|r|\geq r_{\max}} C_p(|r|) \ ,
\end{equation}
which averages the pairing correlation function over the saturation region $|r|\geq r_{\max}$, with $r_{\max}=d_{\max}/2$ (where $d_{\max}$ is the largest distance between two sites of the $L\times L$ cluster) and $\mathcal{M}$ the number of such vectors. We plot $\Delta_{\mathrm{SC}}$ as a function of $1/L$, together with the reference value of Ref.~\cite{shiwei2024science}, obtained with constrained-path auxiliary-field quantum Monte Carlo (AFQMC). The extrapolation to $1/L\to0$ yields a finite value, $\Delta_{\mathrm{SC}}=0.022(1)$, in agreement with the previous estimate within error bars.

\begin{figure}[t]
  \centering
  \includegraphics[width=\columnwidth]{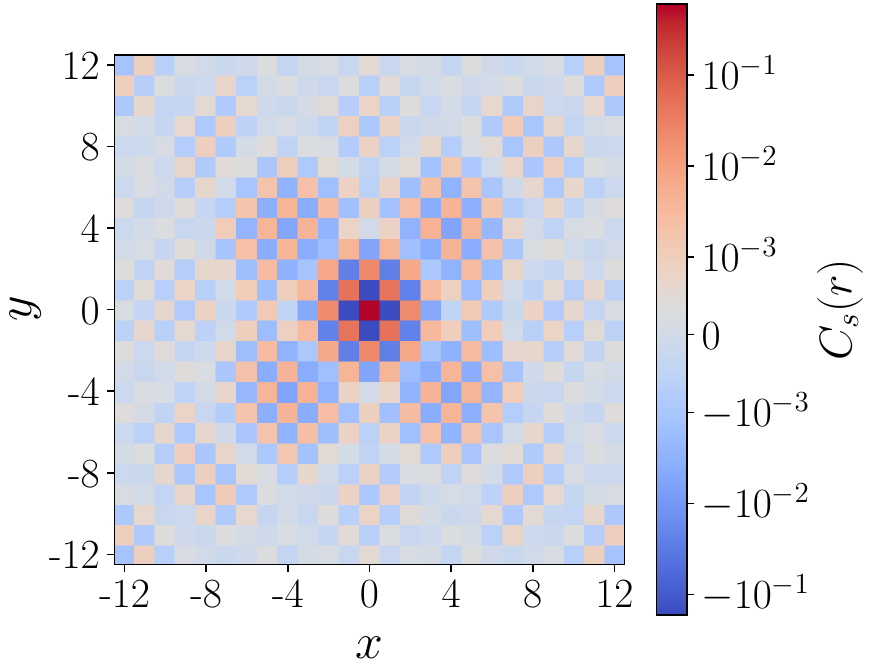}
    \caption{\label{fig:spin_corr} Real space spin-spin correlation function $C_s(r)$ (see definition in the main text), with $r=(x,y)$ the relative positions, on the $24\times24$ lattice for the $t$-$t'$ Hubbard model with periodic boundary conditions, at $t'=-0.2$, $U=8.0$, doping $\delta=1/8$.}
\end{figure}

We note that the result of Ref.~\cite{shiwei2024science} was obtained on cylindrical geometries, with open boundary conditions along the long direction, by introducing an external pairing field~\cite{shiwei2024science, qin2020} and averaging over twisted boundary conditions. Here, instead, we obtain a consistent value by considering only square clusters with periodic boundary conditions and measuring the superconducting order parameter directly from the long-distance behavior of correlation functions. Extracting the thermodynamic properties of the Hubbard model from large square clusters with periodic boundary conditions marks a methodological advance for the numerical study of strongly correlated fermions.

In \cref{fig:spin_corr} we show the spin-spin correlation function $C_{s}(r) = \langle \hat{\boldsymbol{S}}_{0}\cdot \hat{\boldsymbol{S}}_{r} \rangle$ for a $24\times24$ lattice. The correlations follow an antiferromagnetic pattern, modulated by a longer-wavelength stripe envelope~\cite{shiwei2016qmc,roth2025,science2019dmrg}, so that stripe order is clearly present on this cluster. Whether this modulation survives in the thermodynamic limit and long-range stripe order coexists with superconductivity,  or whether these stripe correlations only exist over intermediate scales 
is a computationally delicate issue which should be addressed in future work.
Because the SBP wave function is translationally invariant, the physics of stripes is 
revealed only by the correlation function and not by the one-body observables~\cite{viteritti2026bias}: the local quantity $\langle \hat{S}^{z}_{i} \rangle$ remains uniform across the lattice, as mentioned above, in contrast to the broken-symmetry solutions of \cref{fig:patterns}.

\section{Discussion} In variational optimization, the structure of the energy landscape is determined both by the Hamiltonian and by the functional form of the wave function. As a consequence, the choice of ansatz controls which regions of the variational manifold can be efficiently explored and which minima are most easily reached during training. For the $t$-$t'$ Hubbard model, we have shown that wave functions that explicitly break translational invariance generate a landscape with several competing local minima associated with distinct broken-symmetry pure stripe patterns that typically display weak or no superconducting order~\cite{viteritti2026bias}.

To address this, we introduced the Symmetry-Preserving Backflow Pairing, or SBP, wave function, which enforces translational invariance by construction and removes these symmetry-broken solutions from the accessible variational landscape. Independently of the initialization of the neural-network parameters, the SBP ansatz reaches state-of-the-art variational energies and converges to solutions with consistent physical properties. Moreover, because of its translationally invariant structure, the optimized wave function can be transferred from a given system size to a larger one, leading to a substantial reduction of the computational cost on large systems.

Using the SBP wave function, we computed the ground state of system sizes up to $24\times24$ with $504$ electrons, which were previously out of reach for the $t$-$t'$ Hubbard model using variational neural-network approaches. Our results provide robust evidence that the $t$-$t'$ Hubbard model at $t'=-0.2$, $U=8.0$, and doping $\delta=1/8$ supports superconducting order, thereby resolving a long-standing question about this parameter regime.

A natural direction for future work is to extend these calculations to other dopings and interaction strengths in order to clarify the ground-state phase diagram of the Hubbard model and, in particular, the interplay between superconductivity and stripes~\cite{shiwei2024science}. On the methodological side, a natural extension is to replace the pair-product form of \cref{eq:ansatz}, where pairing is restricted to opposite-spin components, with a more general and expressive Pfaffian parametrization~\cite{chen2025,roth2025,viteritti2026bias}. This would allow the variational state to capture more general pairing structures and could be crucial for obtaining accurate ground states in regions of the phase diagram where different orders are strongly intertwined~\cite{ido2018, darmawan2018}. More broadly, since the SBP ansatz is based on physically motivated principles rather than being tailored to a specific model, we expect it to provide a powerful and flexible tool for the study of strongly correlated fermionic systems, 
such as realistic multi-band models of superconductors involving several 
orbitals per unit cell.

\begin{figure*}[t]
  \centering
  \includegraphics[width=2.1\columnwidth]{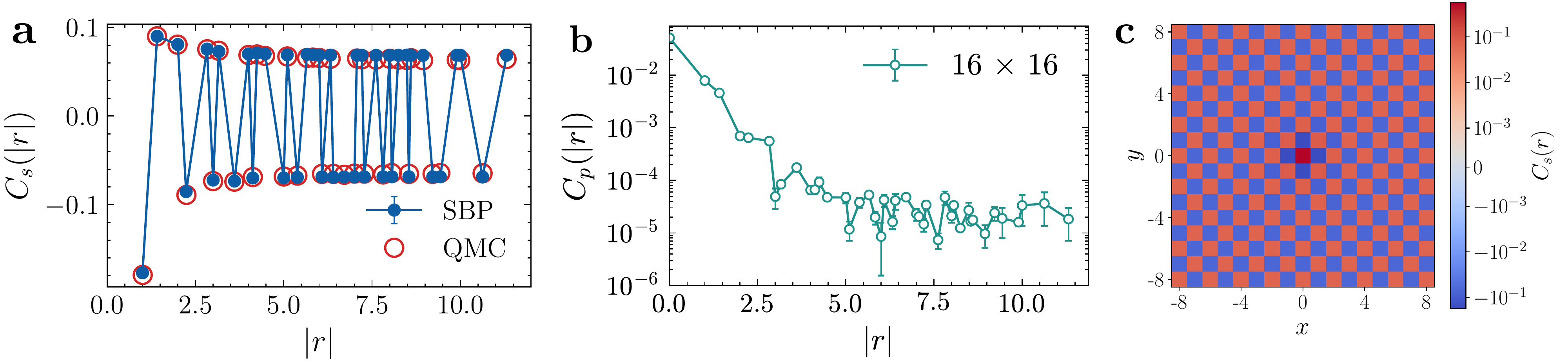}
  \caption{Benchmark of the SBP ansatz against numerically exact quantum Monte Carlo (QMC) for the pure Hubbard model at $t'=0.0$, $U=4.0$, and half filling, on a $16 \times 16$ square lattice with periodic boundary conditions. \textbf{Panel (a)}: spin-spin correlation function $C_s(|r|)$ obtained with the SBP ansatz (filled blue circles) and with QMC (empty red circles), as a function of the distance $|r|$; the QMC data are taken from Ref.~\cite{shiwei2016qmc}. \textbf{Panel (b)}: $d$-wave pairing correlation function $C_p(|r|)$ obtained with the SBP ansatz, as a function of the distance $|r|$. \textbf{Panel (c)}: real-space spin-spin correlation function $C_s(r)$ obtained with the SBP ansatz, with $r=(x,y)$, the color of each point encodes the value of $C_s(r)$ as indicated by the colorbar. \label{fig:fig5}}
\end{figure*}

\section{Methods}
\subsection{Comparison with Quantum Monte Carlo at half filling}
To validate the SBP state, we benchmark it against numerically exact results for the pure Hubbard model $(t'=0.0$, $U=4.0)$ at half filling $(N_e = N)$ on a square lattice with periodic boundary conditions. In this regime, the model is free of the sign problem, so quantum Monte Carlo (QMC) provides exact reference data against which the variational results can be compared~\cite{shiwei2016qmc}. Moreover, this case probes the complementary physical regime to the doped case: at half filling, the ground state exhibits strong antiferromagnetic order and no superconductivity. We compute the spin-spin correlation function $C_{s}(r) = \langle \hat{\boldsymbol{S}}_{0}\cdot \hat{\boldsymbol{S}}_{r} \rangle$ with the SBP ansatz and compare it with the QMC result of Ref.~\cite{shiwei2016qmc} on a $16 \times 16$ lattice. The real-space spin-spin correlation functions $C_s(r)$, shown in \cref{fig:fig5}(c), display the expected antiferromagnetic checkerboard structure, without the longer-wavelength stripe modulation observed in the doped case. As shown in \cref{fig:fig5}(a), the variational results agree with QMC over the full range of distances, demonstrating that the SBP wave function quantitatively captures the antiferromagnetic correlations of the half-filled model. We then compute the $d$-wave pairing correlation function $C_p(r)$ defined above. As shown in panel (b) of \cref{fig:fig5}, in contrast to the doped case, $C_p(r)$ decays rapidly to very low values, compatible with a state with no $d$-wave superconducting order in the thermodynamic limit.

\begin{figure}[t]
  \centering
  \includegraphics[width=\columnwidth]{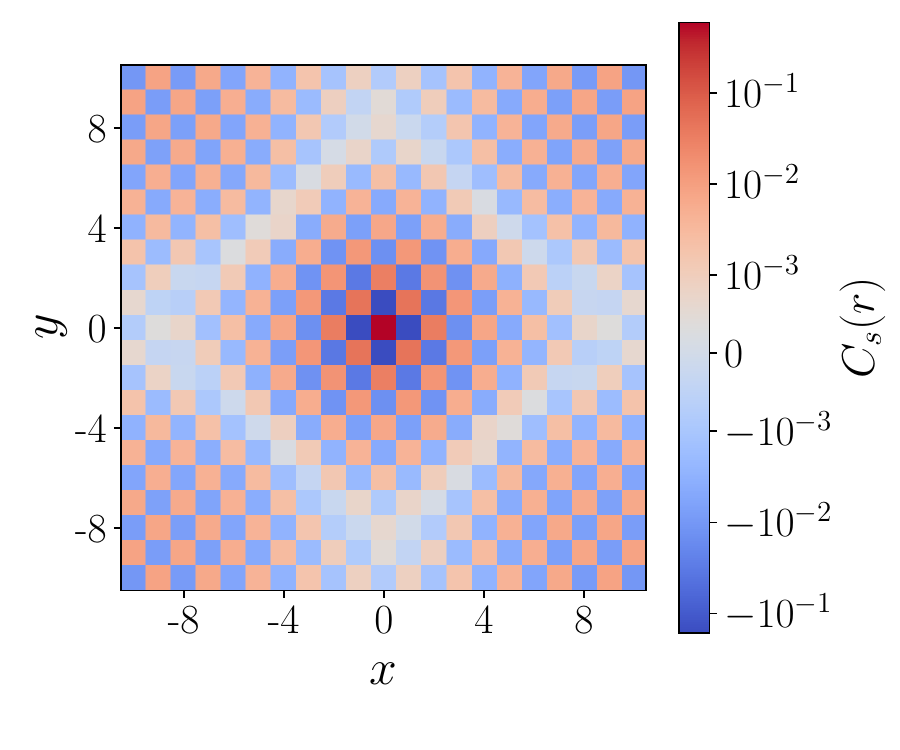}
    \caption{\label{fig:spin_corr_tp0} Real space spin-spin correlation function $C_s(r)$ (see definition in the main text), with $r=(x,y)$ the relative positions, on the $20\times20$ lattice for the pure Hubbard model ($t'=0.0$) with periodic boundary conditions, at $U=8.0$ and doping $\delta=1/8$.}
\end{figure}

\subsection{Spin correlations in the pure Hubbard model at finite doping}

To further assess the ability of the SBP wave function to capture the physics of different parameter regimes, we consider the pure Hubbard model ($t'=0.0$) at $U=8.0$ and doping $\delta=1/8$ on a $20\times 20$ lattice. In Fig.~\ref{fig:spin_corr_tp0}, we show the corresponding real-space spin-spin correlation function $C_s(r)$. As in the $t'=-0.2$ case of Fig.~\ref{fig:spin_corr}, the correlations display an antiferromagnetic pattern modulated by a stripe envelope; however, two clear differences emerge. First, the period of the stripe modulation is longer at $t'=0.0$, in agreement with previous studies at this doping~\cite{shiwei2022prr,zheng2017,science2019dmrg, qin2020}, whereas at $t'=-0.2$ the modulation is shorter, consistent with the partially filled stripes reported 
e.g. in Refs.\cite{shiwei2024science,npj2026,roth2025}.
Second, the magnitude of the spin-spin correlations at large distances is significantly larger at $t'=0$, indicating that magnetic order is more robust in the pure Hubbard model. Both features are consistent with the established picture in which a negative $t'$ frustrates and weakens stripe order, shortening its period and partially melting the magnetic correlations. We stress that since the SBP wave function is translationally invariant, these modulations appear only in the correlation functions, while the one-body observables remain uniform across the lattice.

\subsection{Pairing correlation function}

We define the $d$-wave spin-singlet pair operator consistent with Refs.~\cite{qin2020, shiwei2024science} as
\begin{equation}
\hat{\Delta}_r
=
\frac{1}{4}
\sum_{\eta}
h_{r,\eta}
\frac{\left(
\hat{c}_{r\uparrow}\hat{c}_{r+\eta,\downarrow}
-
\hat{c}_{r\downarrow}\hat{c}_{r+\eta,\uparrow}
\right)}{\sqrt{2}}
 \ ,
\end{equation}
the sum over $\eta$ runs over the four nearest neighbors of $r$, and $h_{r,\eta} = +1$ on horizontal bonds and $h_{r,\eta} = -1$ on vertical bonds, so that the pair carries symmetry $d_{x^2-y^2}$. The corresponding pairing correlation function between the origin and a generic lattice site $r$ is $C(r)
=
\langle
\hat{\Delta}^{\dagger}_0\hat{\Delta}_r
\rangle$.
Expanding the singlet operators gives
\begin{align}
C(r)
=
&
\frac{1}{32}
\sum_{\eta,\eta'}
h_{0,\eta} h_{r,\eta'}
\Big[
\langle
\hat{c}_{\eta,\downarrow}^{\dagger}
\hat{c}_{0,\uparrow}^{\dagger}
\hat{c}_{r,\uparrow}
\hat{c}_{r+\eta',\downarrow}
\rangle
-
\langle
\hat{c}_{\eta,\downarrow}^{\dagger}
\hat{c}_{0,\uparrow}^{\dagger}
\hat{c}_{r,\downarrow}
\hat{c}_{r+\eta',\uparrow}
\rangle
\nonumber\\
&
-
\langle
\hat{c}_{\eta,\uparrow}^{\dagger}
\hat{c}_{0,\downarrow}^{\dagger}
\hat{c}_{r,\uparrow}
\hat{c}_{r+\eta',\downarrow}
\rangle
+
\langle
\hat{c}_{\eta,\uparrow}^{\dagger}
\hat{c}_{0,\downarrow}^{\dagger}
\hat{c}_{r,\downarrow}
\hat{c}_{r+\eta',\uparrow}
\rangle
\Big] \ .
\end{align}
At large separations, spin-singlet superconducting off-diagonal long-range order implies that the four spin-resolved contributions become equivalent once their explicit signs are taken into account. For computational efficiency, we therefore retain only one of them and define
\begin{equation}
C_p(r)
=
\frac{1}{8}
\sum_{\eta,\eta'}
h_{0,\eta} h_{r,\eta'}
\left\langle
\hat{c}_{\eta,\downarrow}^{\dagger}
\hat{c}_{0,\uparrow}^{\dagger}
\hat{c}_{r,\uparrow}
\hat{c}_{r+\eta',\downarrow}
\right\rangle
-
\mathcal{N}_{0,r} \ ,
\end{equation}
where
$
\mathcal{N}_{0,r}
=
\frac{1}{8}
\sum_{\eta,\eta'}
h_{0,\eta} h_{r,\eta'}
\langle
\hat{c}_{\eta,\downarrow}^{\dagger}
\hat{c}_{r+\eta',\downarrow}
\rangle
\langle
\hat{c}_{0,\uparrow}^{\dagger}
\hat{c}_{r,\uparrow}
\rangle$
is the disconnected contribution, retaining only the contraction that conserves both particle number and total $S^z$~\cite{roth2025}. This subtraction does not affect the asymptotic value in the thermodynamic limit. The two definitions need not coincide at finite separation, but with the normalization above $C(r) \rightarrow C_p(r)$ as $|r|\rightarrow\infty$.

\subsection{Transformer architecture and optimization protocol}
The transformer architecture processes the input configurations ${\boldsymbol{n}=(n_{0\uparrow},\ldots,n_{N-1\uparrow},n_{0\downarrow},\ldots,n_{N-1\downarrow})}$ by linearly embedding each local occupation $n^{\sigma}_i$ into a $d$-dimensional vector space, producing the sequence ${(x^{\uparrow}_0, \dots, x^{\uparrow}_{N-1}, x^{\downarrow}_0, \dots, x^{\downarrow}_{N-1})}$ with $x^{\sigma}_i \in \mathbb{R}^d$. This sequence is processed by a stack of layers, each combining a factored self-attention mechanism~\cite{rende2025queries,rende2024prr} with spatial bias~\cite{viteritti2026approaching} and a fully connected network. The resulting output sequence ${(y^{\uparrow}_0, \dots, y^{\uparrow}_{N-1}, y^{\downarrow}_0, \dots, y^{\downarrow}_{N-1})}$, with $y_i^\sigma \in \mathbb{R}^d$, enters the output layer of \cref{eq:backflow} to produce the backflow pairing orbitals of the SBP wave function. 
A detailed description of the transformer architecture and its hyperparameters is given in Refs.~\cite{viteritti2025prb,viteritti2023prl}. Here we use $n_l=8$ layers, $h=12$ heads, and an embedding dimension of $d=72$. To accelerate the optimization of $d$-wave superconducting correlations, we seed the singlet $d_{x^2-y^2}$ pairing channel by adding to $f(\boldsymbol{n})$ [see \cref{eq:backflow}] a configuration-independent matrix $f_d$. Its only nonzero entries connect opposite spins on nearest-neighbor bonds and are defined by $(f_d)_{{r},{r}+{\eta}}= \Delta_d\, h_{{r},{\eta}}$, where $\Delta_d$ is a trainable scalar parameter and $h_{{r},{\eta}}=+1$ ($h_{{r},{\eta}}=-1$) for ${\eta}=\pm\hat{{x}}$ (${\eta}=\pm\hat{{y}}$). We numerically check that this additional term only improves the learning speed; the final optimized results are independent of whether it is included.

The wave function is optimized within the Variational Monte Carlo framework~\cite{becca2017} with Stochastic Reconfiguration~\cite{sorella1998}, using the linear-algebra identity of Refs.~\cite{rende2024stochastic,chen2024empowering} and the MARCH optimizer~\cite{solving2025}. This identity is convenient in our setup, where the number of samples ${M=12288}$ is much smaller than the number of parameters ${P\approx5 \times 10^5}$, as it reduces the size of the linear system to be solved from $P\times P$ to $M\times M$~\cite{rende2024stochastic}.

After the optimization of the SBP wave function [see \cref{eq:ansatz}] we apply quantum number projection~\cite{tahara2008} to restore the rotational $C_{4v}$ symmetry~\cite{viteritti2025prb, rende2024stochastic}. Specifically for the $12 \times 12$ and $16 \times 16$ clusters, we optimize for $2 \times 10^3$ steps after the symmetry restoration; however, for the $20 \times 20$ and $24 \times 24$ clusters, we only perform the sampling without optimization.

\acknowledgments{
We are grateful to Shiwei Zhang, Chris Roth, Steven White, Alessandro Laio for useful discussions. We thanks Federico Becca, Yusuke Nomura, Yahui Zhang, Giuseppe Carleo and Miles Stoudenmire for comments on the draft. The Flatiron Institute is a division of the Simons Foundation, and we acknowledge support of the Flatiron Scientific Computing Core. This work used $7000$ GPU hours on H200 NVIDIA GPUs. LLV is supported by SEFRI under Grant No. MB22.00051 (NEQS - Neural Quantum).
}

\bibliography{refs}

\end{document}